\documentclass[structabstract]{aa}
\usepackage{graphicx}
\usepackage{amsmath}
\usepackage{multirow}
\usepackage{txfonts}
\newcommand{\Msol}{M_\odot}

\newcommand{\ud}{{\mathrm d}}

\begin{document} 
\title{
The MEMO project:
Combining all microlensing surveys to search for intermediate-mass Galactic black holes.
}
\author{
A.~Mirhosseini\inst{1},
M.~Moniez\inst{1}
}
\institute{
Laboratoire de l'Acc\'{e}l\'{e}rateur Lin\'{e}aire,
{\sc IN2P3-CNRS}, Universit\'e de Paris-Sud, B.P. 34, 91898 Orsay Cedex, France
}

\offprints{M. Moniez,\\ \email{ moniez@lal.in2p3.fr}}

\date{Received 27/11/2017, accepted 08/09/2018}
%

\abstract
{}{
The microlensing surveys MACHO, EROS, MOA and OGLE (hereafter called MEMO) have searched for microlensing toward the
Large Magellanic Cloud for a cumulated duration of 27 years.
We study the potential of joining these databases to search for very massive objects,
that produce microlensing events with a duration of several years.
\\
}
{
We identified the overlaps between the different catalogs and compiled their time
coverage to identify common regions where a joint microlensing detection algorithm
can operate. We extrapolated a conservative global microlensing detection efficiency based on simple
hypotheses, and estimated detection rates for multi-year duration events.
\\
}
{
Compared with the individual survey searches, we show that a combined search for long timescale microlensing
should detect about ten more events
caused by $100\Msol$ black holes if these objects have a major contribution to the Milky Way halo.
\\
}
{
Assuming that a common analysis is feasible, meaning that the difficulties that arise from using different passbands
can be overcome, we show that the sensitivity of such an analysis might enable us to quantify the Galactic black hole component.
}

\titlerunning{The MEMO project}
\authorrunning{Mirhosseini, Moniez}

\maketitle

\section{Introduction}

Following Paczy\'nski's seminal publication (\cite{pacz1986}),
several groups have operated survey programs, beginning in 1989,
to search for compact halo objects within the Galactic halo.
The initial challenge for the teams of the Exp\'erience de Recherche d'Objets Sombres (EROS)
and MAssive Compact Halo Objects (MACHO)
was to clarify the status of the missing hadrons in the Milky Way.
Since the first discoveries by EROS (\cite{eroslmc}),
MACHO (\cite{machlmc}), and the
Optical Gravitational Lensing Experiment (OGLE; \cite{oglpr}), thousands of microlensing effects have been
detected in the direction of the Galactic center, as have
a handful of events toward the Galactic spiral arms (\cite{BS7ans}) and very
few events toward the Magellanic Clouds.
\\
In this letter, we focus on the data of the surveys toward the Large Magellanic Cloud (LMC), which is
the statistically dominant target for probing the dark compact objects of the Galactic halo (the others are the Small
Magellanic Cloud, SMC, and M31).
Our purpose is to explore the potential of a combined search for very long duration events that are caused by
intermediate-mass black holes that are now known to exist, thanks
to the recently discovered gravitational waves (\cite{GW1}), and that can be
considered as a possible candidate for the Galactic halo dark matter (\cite{Bird}).

In Sect. \ref{section:surveys} we describe the database compilation from the different surveys.
In Sect. \ref{section:combined_analysis} we list the common catalogs with their time coverage to define the main ingredients of a common analysis.
We propose a conservative approach to extrapolate the efficiency of a global analysis
from the efficiency of the longest duration survey and cadencing information in Sect. \ref{section:efficiency};
based on this efficiency, we estimate in Sect. \ref{section:results} the minimum number of events that may be
detectable in a combined analysis as a function of the deflector mass for a
Milky Way halo completely made of such mass deflectors.
In Sect. \ref{section:discussion} we discuss the feasibility and difficulties of a combined analysis and conclude.
\section{Introduction to the microlensing effect}
\label{section:basics}
When a massive compact object passes close enough to the line of sight of a source, this source is
temporarily magnified. This is called the gravitational microlensing effect.
A review of the microlensing formalism can be found in \cite{Schneider} and \cite{Rahvar}.
When a single point-like lens of mass $M$ located at distance $D_L$ deflects the
light from a single point-like source located at distance $D_S$, the magnification $A(t)$
of the source luminosity as a function of time $t$ is given by (\cite{pacz1986})
\begin{equation}
\label{magnification}
A(t)=\frac{u(t)^2+2}{u(t)\sqrt{u(t)^2+4}}\ ,
\end{equation}
where $u(t)$ is the distance of the lensing object to the undeflected line of sight, divided by
the Einstein radius $R_{\mathrm{E}}$ :
\begin{equation}
R_{\mathrm{E}}\!\! =\!\! \sqrt{\frac{4GM}{c^2}D_S x(1-x)}
\simeq\! 4.54\ \mathrm{AU}.\left[\frac{M}{\Msol}\right]^{\frac{1}{2}}\!
\left[\frac{D_S}{10 kpc}\right]^{\frac{1}{2}}\!\!
\frac{\left[x(1-x)\right]^{\frac{1}{2}}}{0.5}, \nonumber
\end{equation}
$G$ is the Newtonian gravitational constant, and $x = D_L/D_S$.
When the lens that moves at a constant relative transverse velocity $v_T$ reaches its minimum
distance $u_0$ (impact parameter) to the undeflected line of sight
at time $t_0$, $u(t)$ is given by $u(t)=\sqrt{u_0^2+(t-t_0)^2/t_{\mathrm{E}}^2}$,
where $t_{\mathrm{E}}=R_{\mathrm{E}} /v_T$ is the lensing timescale:
\begin{eqnarray}
t_{\mathrm{E}} \sim
79\ \mathrm{days} \times 
\left[\frac{v_T}{100\, km/s}\right]^{-1}
\left[\frac{M}{\Msol}\right]^{\frac{1}{2}}
\left[\frac{D_S}{10\, kpc}\right]^{\frac{1}{2}}
\frac{[x(1-x)]^{\frac{1}{2}}}{0.5}\; . 
\label{eq.tE}
\end{eqnarray}

\subsection{Microlensing event characteristics}
The so-called simple microlensing effect (point-like source and lens
with rectilinear motions) has the following characteristic
features: 
Because an alignment is not likely,
the event should be singular in the history of the source
(as well as in the history of the deflector);
the 
magnification, independent of the color,
is a simple function of time
depending only on ($u_0, t_0, t_{\mathrm{E}}$),
with a symmetrical shape;
as the source and the deflector are independent,
the prior distribution of the events' impact parameters must be uniform;
all stars at the same given distance have the same probability of beeing lensed;
therefore the sample of lensed stars should be representative
of the monitored population at that distance, particularly with respect to
the observed color and magnitude distributions.

This simple microlensing description can be complicated in
many different ways: for example, with multiple lens and source systems (\cite{Mao}),
extended sources (\cite{Yoo}), and parallax effects (\cite{Gould})
because the apparent motion of the lens is non-rectilinear, induced by the orbital motion of Earth.

\subsection{Statistical observables}
The optical depth up to a given source distance, $D_S$, is defined as the
probability of intercepting
a deflector's Einstein disk, which corresponds to a magnification $A > 1.34$.
This is found to be independent of the deflector mass function
\begin{equation}
\tau=\frac{4 \pi G D_S^2}{c^2}\int_0^1 x(1-x)\rho(x) \ud x\, ,
\end{equation}
where $\rho(x)$ is the mass density of deflectors at
distance $x D_S$.

In contrast to the optical depth, the microlensing event durations $t_E$ and consequently
the event rate (deduced from the optical depth and durations) depend on the deflector mass distribution
as well as on the velocity and spatial distributions.

The expected number of events for a microlensing survey is estimated from
\begin{equation}
N_{exp}= \frac{2}{\pi}\times \tau .N_{stars}T_{obs}  \int \frac{\epsilon(t_E)}{t_E}D(t_E)\ud t_E,
\label{event_number}
\end{equation}
where $N_{stars}$, $T_{obs}$, $\epsilon(t_E)$, and $D(t_E)$ are the
number of monitored stars, the survey duration, the detection efficiency, and the
normalized prior $t_E$ distribution of ongoing microlensing events at a given time.


\section{Four main surveys toward the LMC and their results}
\label{section:surveys}
EROS, a mostly French collaboration, started to observe the LMC in the early 1990s. The first phase of this project consisted of two programs. The first program (\textit{EROS 1 plate}) used 290 digitized photographic plates taken at the ESO one-meter Schmidt telescope from 1990 to 1994 (\cite{eros1}). In the second program (1991-1994), a CCD camera was used to observe a small field in the LMC bar.
The fast cadence of this program made it sensitive to low-mass lenses with an event duration from one hour to three days (\cite{eroslmc}). The observations toward the LMC continued in the second phase of the EROS project (\textit{EROS 2}) using the Marly one-meter telescope at ESO, La Silla, equipped with two 0.95 deg$^2$ CCD mosaics (\cite{Tisserand2007}).
\par The MACHO team, a US-Australian collaboration, accumulated data on the LMC from 1992 for nearly six years with a $1.27 m$ telescope and two cameras with four CCDs each, at the Australian Mount Stromlo site (\cite{Alcock2000b}).
\par The OGLE experiment, a collaboration led by the University of Warsaw, started in 1992 and
is now in its fourth phase (\cite{desc-OGLE4}). OGLE-III used a dedicated $1.3m$ telescope with a $64$ Mpixel camera.
Since 2009, OGLE-IV is equipped with a $256$ Mpixel large-field camera.
This collaboration already found an event (OGLE-2005-SMC-1) that can be interpreted as a binary black hole (with non-luminous
components of $7.3\Msol$ and $2.7\Msol$) within the halo (\cite{OGLE-binary}).
\par The MOA team is a collaboration between New Zealand and Japan, operating a dedicated $1.8m$
telescope that is equipped with a $64$ Mpixel camera (\cite{MOAcam3}).
Since the published information on the MOA database toward the LMC is not as complete as the other collaborations,
we were unable to quantify a potential input to a combined analysis.
\par
\begin{figure}[htbp]
\begin{center}
\includegraphics[width=9cm]{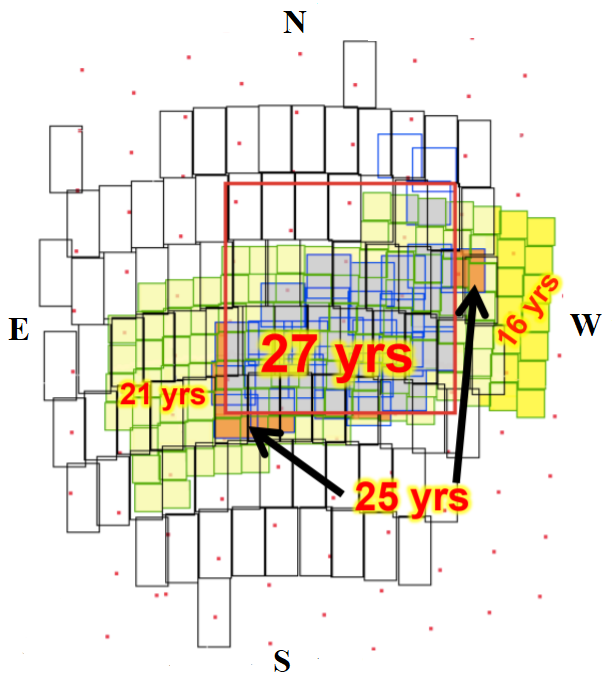}
\caption{Fields monitored toward the LMC: EROS 1 plate (large red square), MACHO fields used to measure the optical depth (\cite{Alcock2000b}, blue squares), EROS-2 (black rectangles), and OGLE-III (green squares). Red dots show the positions of (a fraction) the OGLE-IV field centers. The gray field contains catalog 1, which has been observed for 27 years. The orange and gray field covers catalog 2, which has been observed for 25 years. Catalog 3 is contained in the light yellow, orange, and gray field, and it has been observed for 21 years. Finally, the combination of all the colored fields contains catalog 4, which has been observed for 16 years.
}
\label{fields}
\end{center}
\end{figure}

 Table \ref{surveys} summarizes the EROS, MACHO, OGLE and MOA data taken toward the LMC. Furthermore, the positions of the fields covered by these four surveys toward the LMC are shown in Figure \ref{fields}. Since the main aim of the EROS 1 CCD program was to detect low-mass objects, it has monitored far fewer stars than the other surveys, and we did not use it in the combined analysis considered in the next sections.

\begin{table}
{\scriptsize
\center
\begin{tabular}{c|c|c|c|c|c|c}
      & & T$_{obs}$ & & Field & Objects & \\
     Survey & epoch & (yrs) & Filter & (deg$^2$) & ($\times 10^6$) & Cadence \\
     \hline
     \hline
     EROS 1 & Sep 90- &\multirow{2}{*}{3.5}&R$_E$&\multirow{2}{*}{27.0}&\multirow{2}{*}{4.2}&\multirow{2}{*}{3 days}\\
     plate &Apr 94&&B$_{E}$&&&\\
     \hline
     \multirow{2}{*}{MACHO}&Jul 92- &\multirow{2}{*}{5.7}&R&\multirow{2}{*}{13.4}&\multirow{2}{*}{11.9}&\multirow{2}{*}{4 days}\\
     & Mar 98 &&V&&&\\
     \hline
     \multirow{2}{*}{EROS 2}& Jul 96- &\multirow{2}{*}{6.7}&I&\multirow{2}{*}{84.0}&\multirow{2}{*}{29.2}&\multirow{2}{*}{3 days}\\
     &Feb 03&&V&&&\\
     \hline
      \multirow{2}{*}{OGLE-III}& Sep 01- &\multirow{2}{*}{7.7}&I&\multirow{2}{*}{38.0}&\multirow{2}{*}{19.4}&\multirow{2}{*}{4 days}\\
     & May 09 &&V&&&\\
     \hline
     \multirow{2}{*}{OGLE-IV}& Mar 10- &\multirow{2}{*}{-}&I&\multirow{2}{*}{$>$84}&\multirow{2}{*}{62}&\multirow{2}{*}{4 days}\\
     & present &&V&&&\\
     \hline
    &\multirow{3}{*}{2006-}&\multirow{2}{*}{-}&I&\multirow{3}{*}{31.0}&\multirow{3}{*}{50}&\multirow{3}{*}{1 hour}\\
    MOA2&&&R&&&\\
    cam3&&&V&&&\\
     \hline
\end{tabular}
}
\caption{Observation epochs, durations, filters, LMC field sizes, numbers of monitored objects, and approximate average cadences for each microlensing survey.}
\label{surveys}
\end{table}

\begin{figure}[htbp]
\begin{center}
\includegraphics[width=9cm]{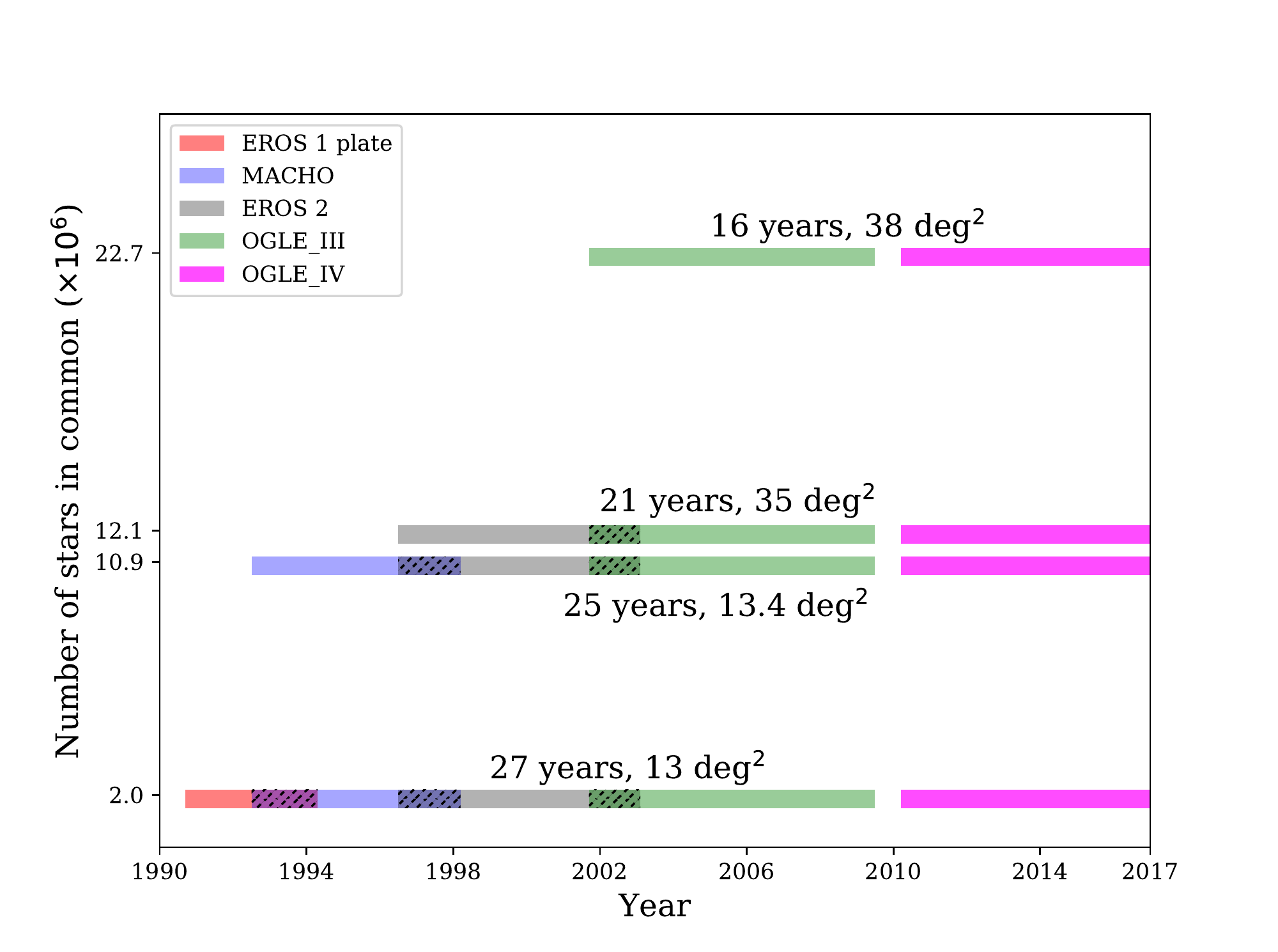}
\caption[]{Catalogs of stars monitored by the different surveys; numbers of stars, field sizes, and observation epochs. Hatched regions show periods in which different surveys overlapped.}
\label{timespan}
\end{center}
\end{figure}

\begin{figure}[htbp]
\begin{center}
\includegraphics[width=8cm]{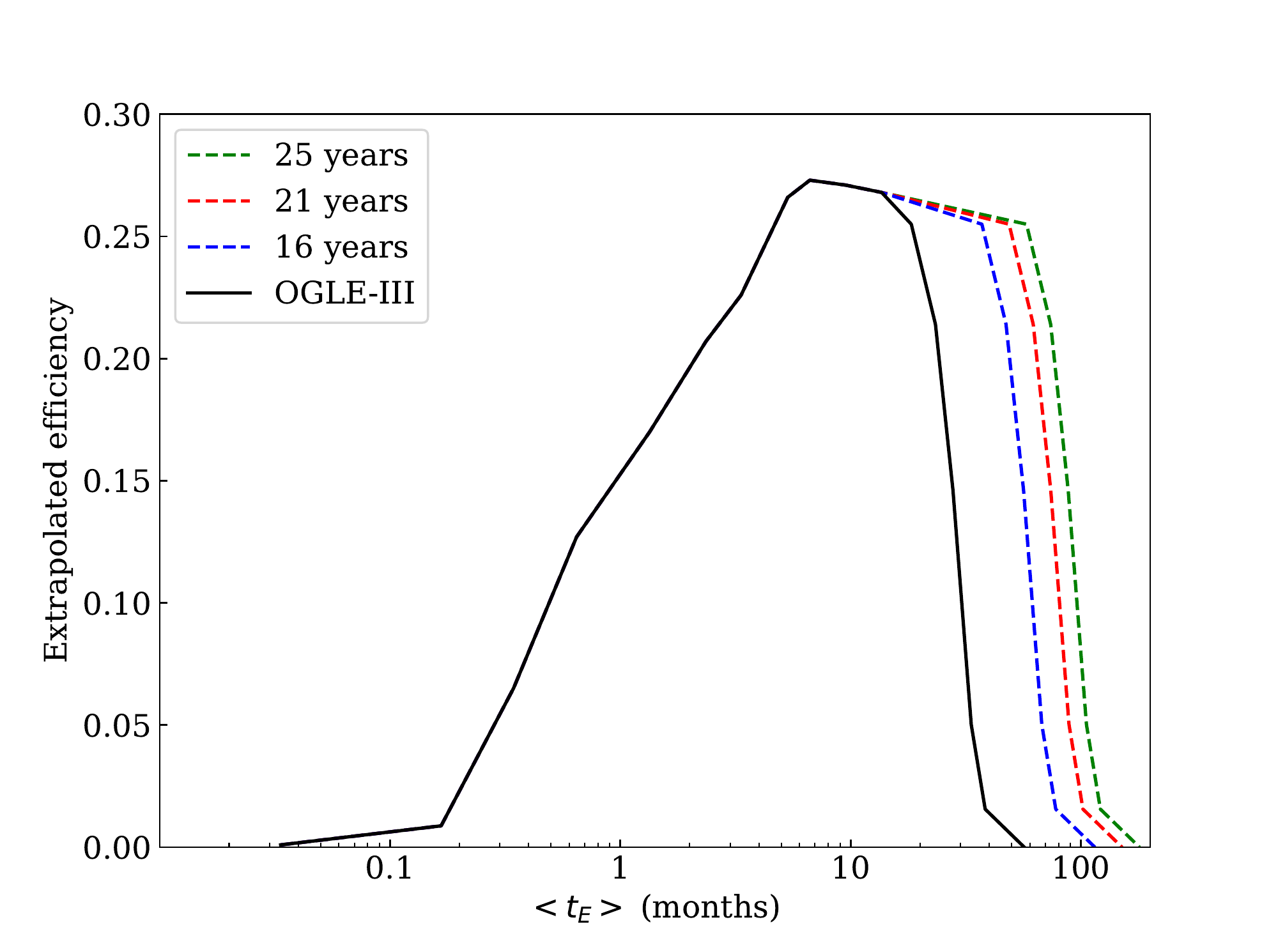}
\caption[]{
Conservative extrapolated microlensing detection efficiencies toward the LMC
for a combined survey analysis as a function of the event characteristic duration $t_E$.}
\label{efficiency}
\end{center}
\end{figure}
\section{Expectations from a simple combined analysis}
\label{section:combined_analysis}
For a combined analysis, we propose to use the full light curve database of all the surveys. We have identified four star catalogs that have been monitored almost continuously for $\Delta T'=$ 16, 21, 25, and 27 years, by at least two surveys toward the LMC (see Fig. \ref{timespan}). The fields covered by these catalogs are shown in Figure \ref{fields}. We estimate below the numbers of expected microlensing events from the populations of each catalog and the corresponding extrapolated average detection efficiencies $\epsilon^{\Delta T'}(t_E)$ of a joint analysis with an overall duration $\Delta T'$.


\subsection{Estimate of a combined analysis efficiency}
\label{section:efficiency}
We have estimated the OGLE-III mean detection efficiency from the 95\% C.L. exclusion diagram published in \cite{OGLE2011} and Eq. (\ref{event_number}).
We find that this mean reconstructed efficiency, estimated for a sample of 19.4 million cataloged objects
and taking into account blending effects and multiple lens impact,
is close to the average published efficiencies for bright stars / dense fields and for all stars / sparse fields.
Since
i) the photometric precisions as a function of the magnitude are similar for all the CCD surveys,
ii) the mean OGLE-III sampling rate is the lowest, and
iii) the mean surface density of the OGLE-III catalog is one of the densest,
we assume that we can conservatively generalize the OGLE-III efficiency for any combination of surveys
covering the same total duration.
The following step aims at estimating microlensing detection efficiencies for combinations of surveys that cover a longer duration.
\par It is well known that the efficiency of all surveys vanishes for long events because their duration is limited, but
we can extrapolate the global efficiency of a combined survey for much longer events from two basic hypotheses:
First, assuming a constant sampling rate, the detection efficiency of a survey for a given Einstein duration increases for longer observations. Second, the efficiency satisfies the following scaling invariance by time dilation by a factor $k$:
\begin{equation}
\epsilon(k\times t_E, LC(k\times t_i))=\epsilon(t_E, LC(t_i)) ,
\end{equation}
where $\epsilon(t_E, LC(t_i))$ is the efficiency of detecting a given microlensing event with an Einstein duration $t_E$ in a light curve $LC$ defined by a series of flux measurements at times $t_i$, and $LC(k\times t_i)$ is the same series of flux measurements, but considered at times $k\times t_i$ instead of $t_i$.
Since the extended light curve obtained by joining surveys should contain more measurements than the simply time-dilated light curve,
these hypotheses allow us to conclude that the detection efficiency for longer events of a combined survey with overall duration $\Delta T'$ should be
\begin{equation}
\epsilon^{\Delta T'}(t_E) \geq \epsilon^{\Delta T}(t_E\times \frac{\Delta T}{\Delta T'}),
\end{equation}
where $\Delta T$ and $\epsilon^{\Delta T}$ are the 
duration and efficiency of the longest individual survey.
We therefore define conservative extrapolated efficiencies as (Fig. \ref{efficiency})
\begin{equation}
\epsilon_c^{\Delta T'}(t_E) = max\left( \epsilon^{\Delta T}(t_E)_{OGLE}\,,\,\epsilon^{\Delta T'}(t_E)_{OGLE}\right).
\end{equation}
The blending has three effects (\cite{blendMACHO-CG}, \cite{Smith2007}, \cite{Tisserand2007}) : it decreases the apparent magnification, it increases the effective number of monitored stars, and
the events appear to have a shorter duration when blending is not taken into account. The two first effects are already taken
into account in the efficiency estimates. The third effect is not estimated, but in our regime, shorter durations mean better detection
efficiency, and we remain conservative by ignoring it.
\subsection{Number of expected microlensing events}
\label{section:results}
To estimate the number of expected events $N_{exp}$, we distinguished between the stellar populations monitored during $\Delta T'=$ 25, 21 and 16 years\footnote{The impact of the EROS1-plate survey was found negligible, because the catalog is smaller and its
marginal contribution to the overall time coverage is very limited (from 25 to 27 years).},
and considered the corresponding extrapolated efficiencies of Figure \ref{efficiency}. Each population number was estimated in its field (Fig. \ref{fields}) from the mean density associated with the shallowest survey:
\begin{equation}
    N_{exp} = \frac{2\tau}{\pi}\times\!\!
    \sum_{\Delta T'\, in\, \{16,21,25\}yrs}\!\! \Delta T' N_{\Delta T'}\int \frac{\epsilon_c^{\Delta T'}(t_E)}{t_E} D(t_E)\ud t_E ,
\end{equation}
where the indices and superscripts of the populations $N$ and efficiencies $\epsilon_c$ denote the combined survey durations.
With this procedure, the expectation for each star is estimated from its most complete light curve obtained by combining all available surveys.

\section{Results and discussion}
\label{section:discussion}
\begin{figure}[htbp]
\begin{center}
\includegraphics[width=9cm]{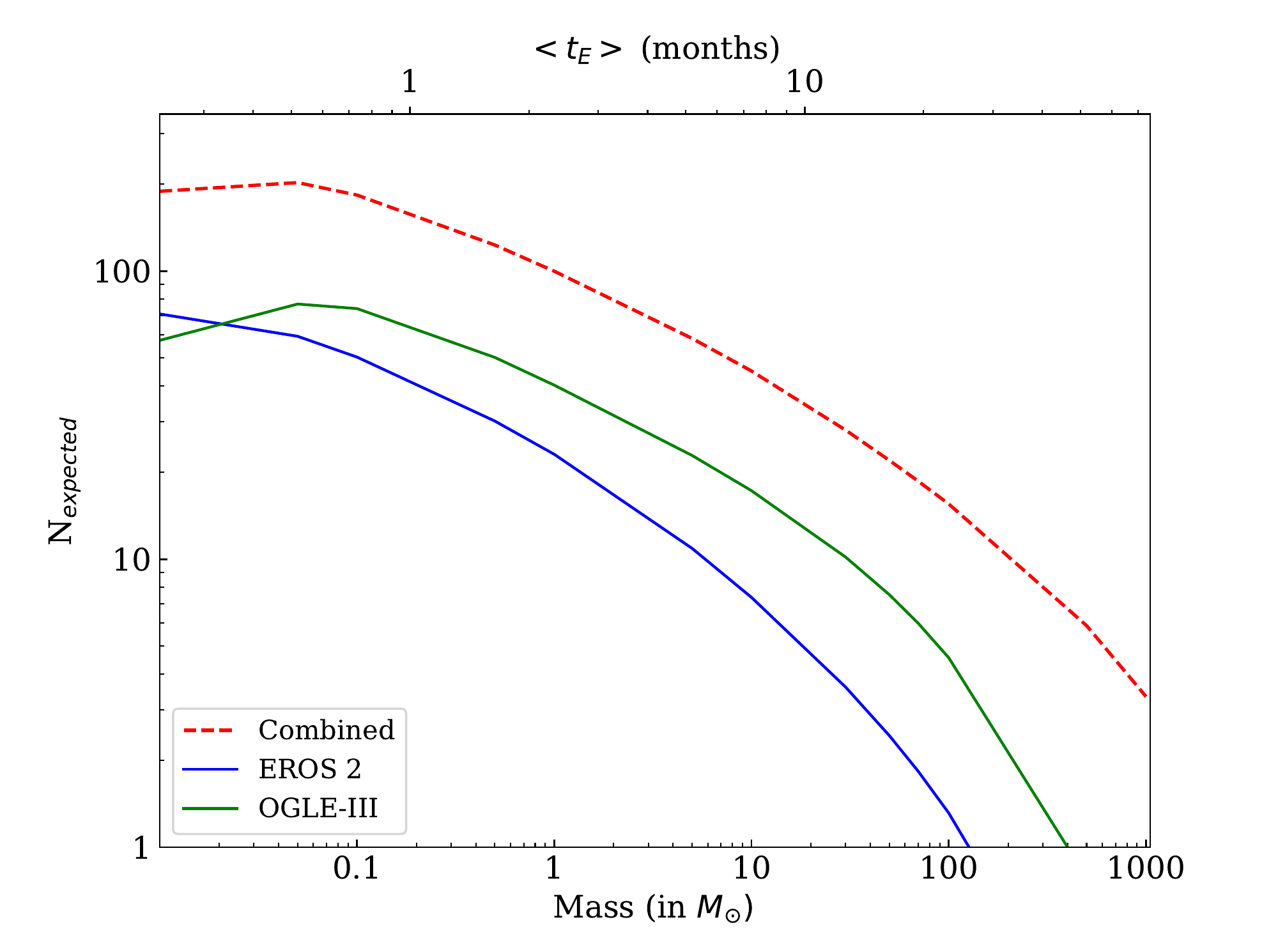}
\caption{Number of expected events for EROS 2 (blue) and OGLE-III (black) and for a combined analysis of the MACHO, EROS and OGLE-III+IV surveys (dashed red line), assuming a standard spherical Galactic halo made of mono-mass (lower abscissa) compact objects. The upper abscissa gives $<t_E>$ corresponding to the deflector mass.
}
\label{expectation}
\end{center}
\end{figure}

The total number of expected events is shown in Fig. \ref{expectation}. The number of events expected for the EROS 2 and OGLE-III surveys alone are shown for comparison. The expectation from a combined survey significantly exceeds the naive addition of the events from each survey. If the standard spherical Galactic halo consists of $100M_{\odot}$ objects, then EROS2 alone would have observed $\sim 1$ event and OGLE-III alone expects $\sim 5$ events, since $\sim 15$ events are expected by combining all the surveys.
This potential encourages us to perform such a joint analysis.
Considering alternative models with different density distributions and kinematics ({\it i.e.}, thick-disk or
non-spherical halo) would need a specific study; but simple first-order scaling can be inferred from Eq. (\ref{eq.tE})
to adjust each curve, and more importantly, the number ratios between the different analysis remains invariant.

Here, the extrapolated efficiencies assume that all events are point sources and point lenses, with rectilinear relative motion, an approximation
that is valid for $90\%$ of the events.
The parallax effect significantly distorts the shape of the light curves only if the Earth orbital velocity ($30 km/s$) is not negligible compared with the lens transverse velocity projected in the solar transverse plane ($\tilde{v} = v_T/(1-x)$).
Toward the LMC, fewer than $\sim 5\%$ of the events that are due to standard halo lenses more massive than $10M_{\odot}$ are expected
to have $\tilde{v} < 5\times 30 km/s$ (\cite{Rahvar2003}), and the light curve distortion is almost never strong enough to significantly affect the detection efficiencies.

The expected complications include the different colour passbands of the surveys, and the inter-seasonal
telescope throughput steps that will need to be considered with particular care to limit misleading  preselections of long-timescale variations.
To this purpose,
the color equations between the different passbands and the zero-points of the instrumental magnitudes will have to be precisely established.
Then a precise global microlensing detection efficiency will need to be computed from specific simulations of combined light curves, taking into account the proper photometric uncertainties and blending of each survey.

A complete interpretation of a candidate sample from which to extract a Galactic halo new component will need precise
modeling of each event that accounts for individual blending and parallax to validate them and/or obtain additional information
on the lens positions. Since we will explore a new time domain,
a previously unknown background of long-timescale variable stars ($>10$ years) could emerge from the
search; in this case, we will have to find a way to distinguish them from microlensing or to statistically subtract them from the sample.
\section{Conclusions and perspectives}
We showed that the joint analysis of the complete available data from all the past, present, and future microlensing surveys toward
the LMC, has the potential of at least doubling the long-timescale event rate expected for the most sensitive survey alone.
Several stages can be foreseen for such a combined analysis:
the first step (corresponding to Fig. \ref{expectation}) uses the already existing light curve catalogs, and a second step would
consist of the complete reanalysis of all the images, searching for variabilities through differential photometry.
Such a combined effort should allow us to establish a census of the black holes within
the Milky Way and help to interpret the rate of gravitational wave detections in the Laser Interferometer Gravitational-Wave
observatory (LIGO) and VIRGO detectors.
 
\end{document}